\documentclass{actasen}
\usepackage{diagbox,multirow,threeparttable}
\newcommand{\upcite}[1]{\textsuperscript{\textsuperscript{\cite{#1}}}}
\usepackage{rotating}
\usepackage{makecell}
\usepackage{subfigure,upgreek,float}

\sfcode`.=1200
\sfcode`:=1200

\allowdisplaybreaks
\usepackage{float,upgreek,colortbl}
\usepackage{url}
\usepackage{makecell}
\usepackage{multicol}
\usepackage{fancyvrb, xcolor}
\usepackage{yhmath}
\usepackage{soul}

\usepackage[numbers,sort&compress]{natbib}
\begin{document}

\setcounter{page}{1}

\Volume{2023}{47}

\runheading{YANG Xu liu et al.}

\title{Simulation Of The Microlensing Effect Near The Critical Curve Of The Galaxy Cluster }

\footnotetext{
\hspace*{3mm}Received 2021--12--17; revised version 2022--01--30

$^{\star}$ A translation of {\it Acta Astron. Sin.~} Vol. 63, No. 6, pp. 66.1--66.13, 2022 \\
\hspace*{5mm}$^{\bigtriangleup}$ luoyu@pmo.ac.cn\\

\noindent 0275-1062/01/\$-see front matter $\copyright$ 2023 Elsevier
Science B. V. All rights reserved 

\noindent PII: }

\enauthor{YANG Xu-liu$^{1, 2}$ \hs\hs CHEN Xue-chun$^{4, 5}$ \hs\hs ZHENG Wen-wen$^{1, 2}$ \hs\hs LUO Yu$^{1, 3}$}{
(\footnotesize 1~{\it Purple Mountain Observatory, Chinese Academy of Sciences, Nanjing 210023})\\
(\footnotesize 2~{\it School of Astronomy and Space Science, University of science and Technology of China, Heifei 230026})\\
(\footnotesize 3~{\it National Basic Discipline Public Science Data Center, Beijing 100190})\\
(\footnotesize 4~{\it Institute for Frontier in Astronomy and Astrophysics, Beijing Normal University, Beijing, China, 102206})\\
(\footnotesize 5~{\it Department of Astronomy, Beijing Normal University, Beijing, China, 100875})\\
}

\abstract{
In the smooth mass distribution model, the critical curve represents a line with magnification divergence on the image plane in a strong gravitational lensing system. Considering the microlensing effects caused by discrete masses, the magnification map in the source plane exhibits a complex structure, which offers a promising way for detecting dark matter.  
However, simulating microlensing near the critical curve poses challenges due to magnification divergence and the substantial computational demands involved.
To achieve the required simulation accuracy, direct inverse ray-shooting would require significant computational resources. 
Therefore we applied a GPU-based code optimized with interpolation method to enable efficient computation on a large scale. 
Using the GPU of NVIDIA Tesla V100S PCIe 32GB, it takes approximately 7000 seconds to calculate the effects of around 13,000 microlenses for a simulation involving $10^{13}$ emitted rays.
Then we generated 80 magnification maps, and select 800 light curves for a statistical analysis of microcaustic density and peak magnification.
}

\maketitle

\section{INTRODUCTION}

Galaxy clusters, as the largest strong lenses, can magnify sources in cosmological distances and produce multiple images. 
 Theoretically, in a galaxy cluster strong lensing system, magnification tends to infinity at a critical curve  in the image plane, which corresponds to a caustic curve in the source plane \cite{B1986}. 
 When a background source crosses from the inside to the outside of the caustic curve, its two images approach the critical curve from both sides, reaching peak magnification near the critical curve, and then merge and disappear \cite{M1991}.

 While a galaxy cluster acts as a strong gravitational lens and generates multiple images of background sources, the presence of compact objects such as stars and black holes that dwell in the galaxy cluster also causes the deflection of the light rays from the background sources, resulting in a microlensing effect. This effect disrupts the expected caustic curve and critical curve predicted by the strong lens model of the cluster. The extremely magnified region transitions from a single line to a diffused band, thereby increasing the probability of observing a magnified high red shift point source.

 Kelly et al \cite{Kelly2016,Kelly2017,K2018} first detected the events of a background star extremely magnified by a galaxy cluster.  
Due to the lensing effect of the galaxy cluster MACSJ1149 at z=0.54,
a blue giant at z=1.49 is magnified by 
over 2000 times.
One image of the source located close to the critical curve of the strong gravitational lensing model of the galaxy cluster.
Its light curve exhibits a significant increase in flux that lasts for several weeks and then fades away, 
shortly after its disappearance, another transient was detected on the other side of the critical curve.
This event can be well explained by the combination of strong lensing and microlensing effects.
The events of extreme magnification of background stars by galaxy clusters has been observed continuously \cite{R2018,C2019,W2022}, which are the consequences of the combined effects of strong lensing by galaxy cluster and microlensing by compact objects within the galaxy cluster.
Typically, the average density of stars within a galaxy cluster is low, and the microlensing effect is not significant.
However, microlensing effect becomes significantly enhanced when image locates in a high magnification region \cite{Dai2021}.
Studying the microlensing effect near critical curve of galaxy cluster provides a valuable way for understanding physical mechanisms behind the extreme magnification of images near the critical curve, which enables us to predict the occurrence and disappearance of these caustic crossing events, furthermore, have a better constrain on the mass fractions of stars and compact dark matter objects \cite{V2017, DK2018, DV2018, O2018}.

The angular separations between multiple images produced by stellar-mass objects are typically in the range of micro-arc seconds \cite{Pac1986,Schneider1987}, making it difficult to resolve the individual micro images through direct observation, only the variation of the total flux can be observed \cite{2001ASPC..239..351W}. 
At the core of microlensing studies lies the simulating of the magnification map in the source plane. The inverse ray-shooting (IRS) method is a direct way for obtaining the magnification map, which mapping a large number of light rays from the image plane back to the source plane by calculating the deflection angle of each ray as they pass through the lens plane \cite{Kayser1986}.

However, the standard IRS method faces two challenges when computing the source plane magnification map near the critical curve: (1) Magnification on critical curve tends to infinity. (2) Microlensing simulation requires a huge amount of computation.
On one hand, the infinite magnification on the critical curve can cause algorithmic failures, and the lens equation needs to be modified accordingly. On the other hand, the required simulation of a large image plane implies the need to trace a large number of light rays, which, with traditional CPU (central processing unit) algorithms, can take months or even years to complete \cite{T2010}.
Based on the IRS method, some more efficient algorithms have been proposed, such as the hierarchical tree method \cite{W1990}, IPM (inverse polygon mapping) algorithm \cite{M2006, M2011}, and GPU (graphics processing units)-based parallel IRS method \cite{T2010,W2021}. 
In particular, the IRS method primarily involves independent accumulation of light ray deflection angles, making it highly suitable for parallel computing using GPU to enable faster computation of the magnification map.

For the extreme magnification events of background stars observed by Kelly et al \cite{Kelly2016,Kelly2017,K2018},  Venumadhav \cite{V2017}  combined analytical and numerical studies in 2017 to investigate the characteristics of microlensing effects near the critical curve. They demonstrated that caustic crossing events are highly sensitive to the surface mass density of microlenses in galaxy clusters and serve as ideal probes for dark matter components.

However, due to computational efficiency limitations, Venumadhav et al \cite{V2017} compute the microlensing effect for low surface mass density to compare with analytical results, then extrapolating their findings to high-density regimes.
In this work, we developed a GPU-based IRS method capable of performing massive calculations. Benefiting from the increased computing speed, we simulated the microlensing effect of a  higher stellar surface mass density near the critical curve, and investigated its statistical characteristics.  
We organized the paper as follow, the description of the lens equation near the critical curve is introduced in section 2. In section 3 we introduce the methods of microlensing simulation near the critical curve from two aspects: parameter settings and ray tracing. In section 4 we introduce the statistical characteristics of the microlensing effect near the critical curve.
A short summary of this work is provided in section 5.

\section{THE LENS EQUATION NEAR THE CRITICAL CURVE}

We consider the smooth lens model under the assumptions of geometric optics and thin lens approximation. $\bm \theta$ denotes the angular position on the image plane, which is mapped to the source plane position $\bm \beta$ through lens equation $\bm \beta=\bm \theta - \bm \alpha \left(\bm \theta\right)$, $\bm \alpha$ represnts the defelection angle. The Jacobian matrix of this mapping is described as
\begin{equation}
\label{eq:A}
    \bm{A}(\bm{\theta}) \equiv \frac{\partial {\bm \beta}(\bm{\theta})}{\partial \bm{\theta}}=
    \left(\begin{array}{cc}
    1-\kappa(\bm{\theta})-\gamma(\bm{\theta}) & 0 \\
    0 & 1-\kappa(\bm{\theta})+\gamma(\bm{\theta})
    \end{array}\right)\ ,
\end{equation}

Eq.(\ref{eq:A}) is a representation of the strong lensing regime in the context of microlensing research, where $\kappa(\bm\theta)$ is the convergence, a dimensionless 
 quantity that describes the mass surface density at the image position, normalized by $\Sigma_{\mathrm{crit}}$.
 Here, $\Sigma_{\mathrm{crit}}=\frac{c^2}{4\pi G}\frac{D_\mathrm{s}}{D_\mathrm{l}D_\mathrm{ls}}$ is the critical mass surface mass density, $c$ denotes the speed of the light, while $G$ denotes the gravitational constant, $D_\mathrm{l s}$, $D_\mathrm{l}$ and $D_\mathrm{s}$ represents the angular diameter distance between the lens and the source, the lens and the observer, the source and the observer,respectively.
 The variable $\gamma(\bm\theta)$ represents shear, which describes the distortion effect on the image. The magnification is the reciprocal of the determinant of the Jacobian matrix, i.e. $\mu=1 / \operatorname{det} \mathbf{A}$

 For every point on the image plane, there exists a unique corresponding point on the source plane, but for a given position on the source plane, there can be multiple corresponding images.
When the source crosses a caustic curve, the images appear or disappear in pairs on the critical curve of the image plane.
In general, the mapping in equation (\ref{eq:A}) is locally invertible, i.e. $\rm{d}\bm{\theta}=\mathbf{A}^{-1}\left(\bm{\theta}\right)\rm{d}\bm{\beta}$. On the critical curve, however, one can find $\operatorname{det}\mathbf{A}=0$.
Thus, we expand convergence and shear up to the first order near the critical curve:
\begin{equation}
    \kappa(\bm{\theta})=\kappa_{0}+\mathbf{\bm\theta} \cdot(\boldsymbol{\nabla} \kappa)_{0}\ , 
\end{equation}
\begin{equation}
    \gamma(\bm{\theta})=\gamma_{0}+\mathbf{\bm\theta} \cdot(\boldsymbol{\nabla} \gamma)_{0}\ , 
\end{equation}

Here, the subscript `` 0 '' indicates quantities on the critical curve. 
In this work, we only consider the fold caustic, that is, on the critical curve, we have $1-\kappa_0-\gamma_0=0$, so the Jacobian matrix near the critical curve is
\begin{equation}
\label{eq:A near cc}
    \mathbf{A}(\bm{\theta})=\left(\begin{array}{cc}
    \mathbf{\bm \theta} \cdot \bm{d} & 0 \\
    0 & 2\left(1-\kappa_{0}\right)
    \end{array}\right)\ , 
\end{equation}

where $\bm{d} \equiv -(\bm\nabla \kappa)_{0}-(\bm\nabla \gamma)_{0}$.
In a local region, the critical curve can be approximated as a straight line, and the direction of $\bm d$ is perpendicular to the critical curve, i.e., $\bm d \cdot \bm \theta=0$. 
Therefore, the lens equation near the critical curve is obtained by integrating Eq(\ref{eq:A near cc}):
\begin{equation}
\label{eq:Lens equation near cc}
    \left\{\begin{array}{l}
    \beta_{x} = \frac{1}{2} d_{x} \theta_{{x}}^{2} + d_{y} \theta_{x} \theta_{y}\ , \\
    \beta_{y} = 2(1-k_0)\theta_{y}\ , \\
    \end{array}\right.
\end{equation}
the subscript `` $x$ '' and `` $y$'' denotes the two dimensions of the  image plane and the source plane.

Now we consider the case where a portion of matter exists in a compact form within the smooth lens.
We add a fraction of microlenses to the smooth lens model and correspondingly subtract a `` mass sheet '' with the same mass density as the microlens mass density $\kappa_{\star}$, ensuring that the total matter mass density remains unchanged.
Thus, the lens equation Eq (\ref{eq:Lens equation near cc}) can be rewritten as:
\begin{equation}
\label{eq:Lens equation near critical plus microlensing}
\bm \beta=\left(\begin{array}{cc}
                \frac{1}{2}d_{x}\theta_{x} + d_{y}\theta_{y} & 0 \\
                0 & 2\left(1-\kappa_{0}\right)
               \end{array}\right) \bm\theta
           - \sum_{k=1}^{N_{\star}} m_{k} \frac{\left(\boldsymbol{\bm \theta}-\boldsymbol{\bm \theta}_{k}\right)}{\left|\boldsymbol{\bm \theta}-\boldsymbol{\bm \theta}_{k}\right|^{2}} - \bm\alpha_{-\kappa_{\star}}\ ,
\end{equation}

the mass and position of the $k$th microlens are denoted as $m_k$ and $\bm{\theta}_k$ respectively. $N_{\star}$ is the  number of the microlenses.
The first term on the right-hand side of Equation (\ref{eq:Lens equation near critical plus microlensing}) represents the ideal mapping relationship between the image plane and the source plane in the smooth model. The values of $\kappa_{0}$ and $\bm{d}$ can be obtained from the strong lensing model.
The second term represents the deflection of light rays contributed from individual microlenses.
The third term represents the deflection caused by the ``negative mass sheet" with a surface mass density of $-\kappa_{\star}$. 
For simplicity, we assume a uniform mass distribution for the microlenses, where each microlens has a mass of $M$, i.e. $m_{k}=1$, and all angles are normalized by $\theta_\mathrm{E}$. Here, $\theta_{\mathrm{E}}$ denotes the Einstein radius of the microlenses:
\begin{equation}
    \label{eq:Einstein angle}
    \theta_{\rm E}=\sqrt{\frac{4GM}{c^{2}} \frac{D_\mathrm{l s}}{D_\mathrm{l} D_\mathrm{s}}}\ .
\end{equation} 

In ray tracing, the computational cost is proportional to the product of the number of microlenses and the total number of light rays. As we can see in section 4, this number is enormous.
Since the deflection angle contributed by each microlens to a light ray is a linear combination of the deflection angles of all microlenses, and each light ray undergoes the same operations independently, without interaction between each other. 
This makes it highly suitable for parallelization on GPUs to improve computational speed. 
Therefore, we employ GPU parallelization to our microlensing simulation  near the critical curve.

\section{SIMULATION NEAR THE CRITICAL CURVE}
In this section, we will introduce the simulation algorithm which is the same as the approach proposed by Zheng et al \cite{Zheng2022}, with the main difference lying in the modification of the lens equation.
The algorithm make full use of the parallel advantages of the GPU to calculate the deflection of light rays simultaneously. Additionally, drawing inspiration from Wambsganss et al \cite{W1990}, different treatments are adopted for microlenses at different distances when calculating the deflection angle: the deflection angle contributed by microlenses in close proximity to the light ray is directly computed, while the deflection angle contributed by microlenses far from the light ray is interpolated. Below, we provide a brief introduction to this algorithm.

\subsection{THE PARAMETERS}
In the angular coordinate system, consider a source plane (referred to as $\rm s$) of size $L_{\mathrm{sx}}\times L_{\mathrm{sy}}$. 
The gravitational scattering of microlenses in the lensing galaxy causes the emitted light rays from the image plane to not fully fall within the $\rm s$ region. This leads to inaccurate magnification calculations near the boundaries \cite{W1990}. 
To ensure the accuracy,  a protection area should be added outside the target source plane $\rm s$ \cite{Katz1986} \cite{Zheng2022}, resulting in an extended region called $\rm s^{\prime}$. The width of the chosen protection zone is set to be $10\sqrt{\kappa_{\star}}$ to achieve approximately 98$\%$ accuracy in the magnification calculation for the s region \cite{Katz1986}\cite{Zheng2022}.
Then, $\rm s^{\prime}=L_{\mathrm{sx}}^{\prime}\times L_{\mathrm{sy}}^{\prime}$:
\begin{equation*}
    L_{\mathrm{sx}}^{\prime}=L_{\mathrm{sx}}+20\sqrt{\kappa_{\star}}\ , L_{\mathrm{sy}}^{\prime}=L_{\mathrm{sy}}+20\sqrt{\kappa_{\star}}\ .
\end{equation*}

From the lens equation, one can obtain the image plane, which is a rectangular region of size $L_{\mathrm{ix}} \times L_{\mathrm{iy}}$. 
Microlenses of in the image plane can be calculated as
\begin{eqnarray*}
    N_{\star}&=&\frac{L_{\mathrm{ix}}
    L_{\mathrm{iy}} \theta_{\mathrm{E}}^2 
    D_{\rm l}^2 \kappa_{\star} \Sigma_{\mathrm{crit}}}{M}\ , \\
    &=&\frac{L_{\mathrm{ix}} L_{\mathrm{iy}}  \kappa_{\star}}{\pi}\ . 
\end{eqnarray*}

\subsection{RAY TRACING}
Assuming that the number of pixels required for the magnification map in the source plane is $N_{\mathrm{pix}}$ and the desired precision is $N_{\mathrm{av}}$ (where $N_{\mathrm{av}}$ represents the average number of rays per grid cell in the magnification map without microlenses), the number of rays to be emitted on the image plane is given by
\begin{equation}
    N_{\mathrm{rays}}=N_{\mathrm{av}}\times N_{\mathrm{pix}}\ .
\end{equation}
The deflection angle produced by each microlens for each ray can be directly calculated as $\bm \alpha_{\star}=\sum_{k=1}^{N_{\star}} m_{k} \frac{\left(\boldsymbol{\bm \theta}-\boldsymbol{\bm \theta}{k}\right)}{\left|\boldsymbol{\bm \theta}-\boldsymbol{\bm \theta}{k}\right|^{2}}$. Even with parallel computation using GPU, the computational workload for this calculation is still large.

Thus, the tree-level algorithm proposed by Wambsganss et al \cite{W1990} is refered to improve efficiency. Based on the distance between microlenses and rays, the microlenses are divided into near microlenses and far microlenses. Consequently, the deflection angle produced by microlenses for a ray is given by $\bm \alpha_{\star}=\bm \alpha_\mathrm{near}+\bm \alpha_\mathrm{far}$. Therefore, the lens equation near the critical curve needs to be modified from Equation (\ref{eq:Lens equation near critical plus microlensing}) as follows:
\begin{equation}
\label{eq:Modified lens equation near critical plus microlensing}
\bm \beta=\left(\begin{array}{cc}
                \frac{1}{2}d_{x}\theta_{x} + d_{y}\theta_{y} & 0 \\
                0 & 2\left(1-\kappa_{0}\right)
               \end{array}\right) \bm \theta
               -\boldsymbol{\alpha}_\mathrm{near}-\boldsymbol{\alpha}_\mathrm{far}-\boldsymbol{\alpha}_\mathrm{-\kappa_{\star}}\ ,
\end{equation}
The deflection angle contributed by near microlenses is directly calculated as $\bm \alpha_{\mathrm{near}}=\sum_{k=1}^{N_\mathrm{near}} m_{k} \frac{\left(\boldsymbol{\bm \theta}-\boldsymbol{\bm \theta}{k}\right)}{\left|\boldsymbol{\bm \theta}-\boldsymbol{\bm \theta}{k}\right|^{2}}$. Here, $N_\mathrm{near}$ represents the number of near microlenses for a given ray. Since the influence of far microlenses on the deflection of the light ray is much smaller compared to near microlenses, and their impact decreases as the distance increases, all the far microlenses can be approximated using interpolation. It ensures both accuracy in the calculation of the deflection angles and improves computational speed.

\begin{figure}[H]
\center{\includegraphics [scale=0.4]{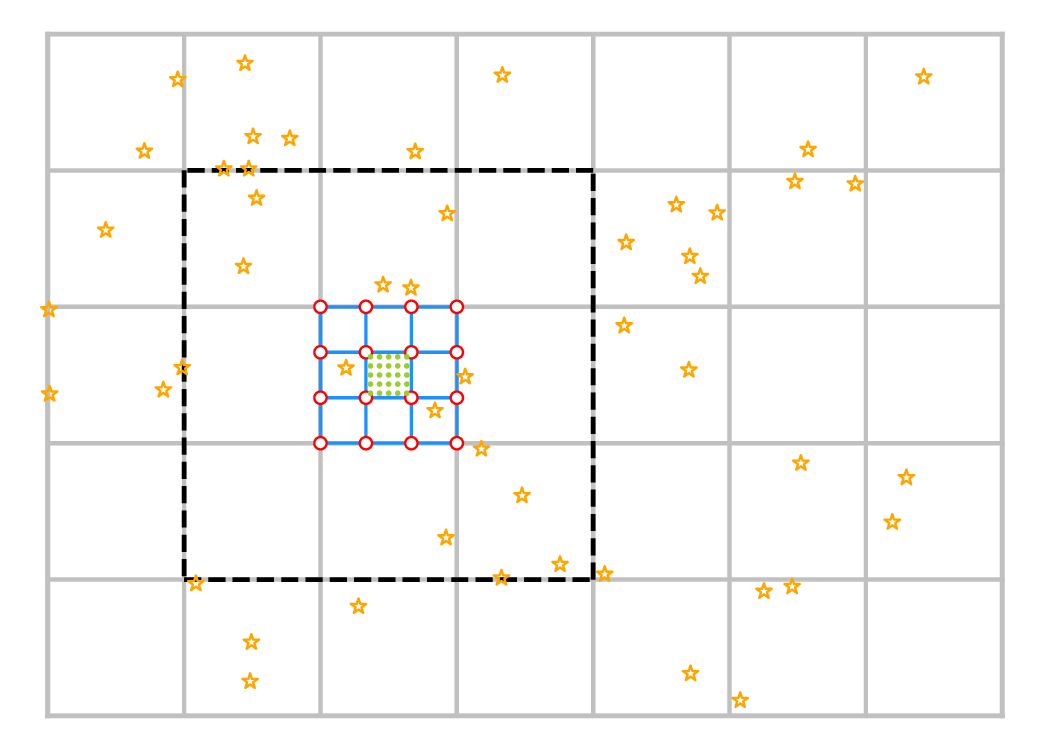}}
\caption{This figure is the three-level of grids sketch map of a image plane. The gray grids represent the level1 grids, the blue grids represent the level2 grids within a level1 grid, the green dots represent the rays within a level2 grid (i.e. the center position of level3 pixels within a level2 grid), and the yellow asterisks represent the microlenses. For the green dots in the figure, the asterisks within the black dotted box represent the near microlenses and the asterisks outside represent the far microlenses. }
\label{fig: three-levels}
\end{figure}

To implement the above method, a three-level grid is set up on the image plane (as shown in Figure \ref{fig: three-levels}), similar to Fig. 3 in the work of Zheng et al \cite{Zheng2022}. The following is a description of the three-level grid.

\subsubsection{Level1: Seting Image Plane}
$N_\mathrm{1x} \times N_\mathrm{1y}$ level 1 grids are setted on the lens plane to distinguish between far and near microlenses. The side length of each level 1 grid is:
\begin{equation}
\label{eq:Size of level1}
    L_\mathrm{1x}=\mathrm{Min}\left[L_{0},\frac{L_\mathrm{ix}}{10}\right]\ , 
\end{equation}
Where $L_{0}=\sqrt{\frac{L_\mathrm{ix} \cdot L_\mathrm{iy}}{N_{\star}}}$, represents the average side length of each microlens in the image plane. When the number of microlenses is large, setting $L_\mathrm{1x}=L_{0}$ ensures that the number of level 1 grids is comparable to the number of microlenses. When the number of microlenses is small, setting $L_\mathrm{1x}=L_\mathrm{ix}/10$ to ensure that the number of level 1 grids is not less than $10 \times 10$.

\subsubsection{Level2: Setting Interpolation Field}
$N_\mathrm{2x} \times N_\mathrm{2y}$ level 2 grids are setted on the lens plane to record the deflection angle information of far microlenses. 
The side lengths of the level 2 grids, $L_\mathrm{2x}$ and $L_\mathrm{2y}$, are set as $L_\mathrm{2x}=L_\mathrm{1x}/20$ and $L_\mathrm{2y}=L_\mathrm{1y}/20$, respectively.
For a given ray, eight fourth-order Taylor coefficients (contributed only by the far microlenses) of the level-2 grid points are computed where the ray is located \cite{W1990}, which is the key procedure for calculating the deflection angle $\bm \alpha_\mathrm{far}$ contributed by the far microlenses.

\subsubsection{Level3: Mapping Light Rays}
The number of level3 grids in the image plane, i.e.total number of rays, is $N_\mathrm{3x} \times N_\mathrm{3y}$. The center position of each pixel corresponds to the position of a ray, and the side length of each pixel is:
\begin{equation}
\label{eq:Size of level3}
    L_\mathrm{3x}=\sqrt{\frac{L_\mathrm{ix} \cdot L_\mathrm{iy}}{N_\mathrm{rays}}}\ . 
\end{equation}
Each ray is assigned to a GPU thread to calculate the deflection angle corresponding to the lens equation (\ref{eq:Modified lens equation near critical plus microlensing}), from both the near lenses and the far lenses, where the contribution by far lenses is calculated by the interpolation procedure mentioned before.

 In this way, the position of each ray in the source plane is obtained. Then the number of rays in each source pixel is calculated and transferred to the CPU, and the magnification  of each pixel is obtained using
\begin{equation}
\label{eq:muij}
    \mu_{ij}=N_{i j} \cdot \frac{S_\mathrm{I}}{S_\mathrm{S}}\ , 
\end{equation}
where $i$ and $j$ represent the horizontal and vertical indices of source pixels in the magnification map, $\mu_{ij}$ and $N_{ij}$ represent the magnification and number of rays for the $(i, j)$-th pixel on the source plane, and $S_\mathrm{I}$ and $S_\mathrm{S}$ represent the pixel areas of the level 3 grid and the source plane, respectively.

The use of GPU and interpolation method allows this algorithm to be two orders of magnitude faster than the traditional GPU parallel IRS algorithm in computing speed. Additionally, the calculation error introduced by interpolation approximation in the deflection angle is approximately on the order of $10^{-7} \theta_\mathrm{E}$ \cite{Zheng2022}.
With it, we are able to handle the substantial computational workload in high magnification regions near caustics.

\section{STATISTICAL CHARACTERISTICS OF MICROLENSING NEAR THE CRITICAL CURVE}

Venumadhav et al \cite{V2017} calculated the microlensing effects in low-density microlens fields ($\kappa_{\star}=6.5\times10^{-5}$ and $\kappa_{\star}=3.25\times10^{-4}$) and extrapolated the results to the case of real microlens mass densities. 
Thanks to the the improvement of high-precision GPU algorithms, we are able to set a higher microlens mass density ($\kappa_{\star}=0.001$) than that used in their study, in order to study the statistical characteristics of microlensing effect near the critical curve.

\subsection{MAGNIFICATION MAPS AND LIGHT CURVES}

We compute the magnification map on the source plane within a range of $200\times 40{\theta^2_{E}}$ near the caustic curve, with a high precision of $N_\mathrm{av}=1000$ and a high resolution of $N_{\mathrm{pix}}=363635 \times 73877$.
Table 1 presents the parameter settings used in the simulation, where $\kappa_{0}$, $\kappa_{\star}$, and $\bm d$ were obtained by Kawamat et al \cite{K2016} when constructing the strong lensing model for the galaxy cluster MACSJ1149.5+2223.

$\kappa_\mathrm{c}=(\theta_{\rm E}{|\bm d|})^{2/3}$ represents a threshold for the surface mass density of stars. If the surface mass density is below this threshold, the caustic and critical curve in the original smooth model are approximately maintained, indicating that the microlensing effect can be neglected. Conversely, if the surface mass density exceeds this threshold, the microlensing effect  becomes significant.
In our calculations, $\kappa_{\star} \gg \kappa_\mathrm{c}$, where the micro critical curves are strongly coupled to the critical curves of the galaxy cluster, resulting in complex critical curve belt \cite{V2017}.

\begin{table}[H]
\tabcolsep=40pt
\center
\caption{Parameter setting in the simulation}
\begin{threeparttable}
\begin{tabular}{cccc}
\hline
Parameter                             & Value(s) \\
\hline
$\kappa_{0}$                     & 0.83 \\
$\bm d$ $\left(\mathrm{arcmin}^{-1}\right)$          & (3.62, -3.41) \\
$\theta_{\rm E}$ $\left(\mathrm{arcsec}\right)$          & $10^{-6}$ \\
$\kappa_\mathrm{c}$\tnote{1}     & $1.9 \times 10^{-5}$ \\
$\kappa_{\star}$                 & 0.001 \\
$N_{\star}$                      & 13745 \\
$N_\mathrm{av}$                  & 1000 \\
\hline
\end{tabular}
\begin{tablenotes}
\item[1] $\kappa_\mathrm{c}$ is the threshold of mass density of star surface, when $\kappa_{\star} \gg \kappa_\mathrm{c}$, the microcritical curves of the microlens are highly coupled with the critical curve of the galaxy cluster, forming the band of corrugated microcritical curves\upcite{V2017}. 
\end{tablenotes}
\end{threeparttable}
\label{tab1}
\end{table}

\begin{figure}[H]
\center{\includegraphics [scale=0.6]{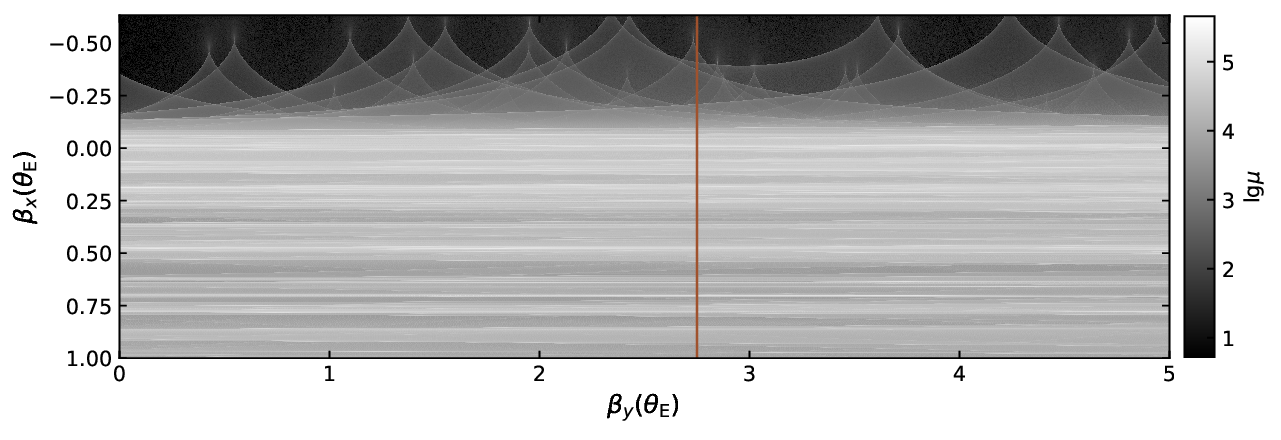}}
\caption{This figure is magnification map of the source plane. The figure shows the region on the source plane in the range of $\beta_{y}\in(0, 5)\theta_{\rm E}$ and $\beta_{x}\in(-0.632456, 1)\theta_{\rm E}$. This region is divided into $9091 \times 2969$ pixels. Parameter settings of the lens field are shown in Table 1. The brown solid line marks part of the trajectory of the background source moving vertically downward along column 5000 of the source plane. The magnification map is shown in logarithmic form, i.e. $\mathrm{log}_{10}\mu$, with different gray levels representing different magnifications. }
\label{fig:sourcemag}
\end{figure}

Then, we shoot about $10^{13}$ light rays from image plane to source plane use the method described in Section 3.2, obtaining the magnification map shown in Fig.\ref{fig:sourcemag}. For convenience of presentation, Fig.\ref{fig:sourcemag} shows only a part of this magnification map, with the range of the two dimensions as $\beta_{y}\in(0,5)\theta_{\rm E}$ and $\beta_{x}\in(-0.632456, 1)\theta_{\rm E}$.
In the smooth model, the caustic corresponds to the horizontal line at $\beta_{x}=0$ in the magnification map, and the magnification below this region decreases according to $1/\sqrt{\beta_{x}}$.
When considering the microlensing effect, it can be seen from  Fig.\ref{fig:sourcemag}, there are clear micro-caustic structures near the caustic line, i.e. micro-caustics overlapping each other, disturb the caustic curve into a micro-caustic band.  
The width of the micro-caustic band is analytically estimated by Venumadhav et al \cite{V2017} :

\begin{equation}
    s_{\rm w} \simeq \frac{2 \theta_{\rm E}}{|\sin a|} \kappa_{\rm c}^{1 / 2}\left(\frac{\kappa_{\star}}{\kappa_{\rm c}}\right)^{2}\ , 
\end{equation}
the meaning of $\kappa_{\rm c}$ is described in Table \ref{tab1}.$a$ represents the angle between the critical curve and the coordinate axes, $\mathrm{tan}a=\left.d_{x}\middle / d_{y}\right.$.

The light curve refers to the variation in magnification of background sources as their positions on the source plane change under the influence of strong lensing by the galaxy cluster and the microlensing  of compact objects within the cluster.
The brown solid line in Fig.\ref{fig:sourcemag} marks a trajectory where the background source moves vertically downwards along a column on the source plane ($i=5000$).
Fig.\ref{fig:lightcurve} presents a light curve and an example of peak identification, panel ${\rm (a)}$ shows the variation of the magnification within each pixel on the source plane $\mu_i$ (solid brown line), and the average magnification $\left \langle \mu \right \rangle$ (dashed black line) as the background source moves along the trajectory. 
It can be seen that the magnification fluctuates around the average magnification, and when the source approaches the caustic curve, both the average magnification and peak magnification show an increasing trend.

\subsection{STATISTICAL CHARACTERISTICS OF PEAKS}

When other parameters are fixed, the number density of micro-caustics reflects the number density of microlenses, while the peak magnification on the light curve reflects the mass of the microlenses. 
Therefore, these two quantities are the most direct indicators for resolving microlensing objects and constraining their mass and mass fraction. 
Venumadhav et al \cite{V2017} analytically estimated that at a distance of one $s_{\rm f}$ from the strong lensing caustic curve in the source plane, 
\begin{equation*}
s_{\rm f}=\left[\ln \left(3.05 N_{\mathrm{\star}}^{1 / 2}\right)\right]^{1 / 2} \theta_{\rm E} \kappa_{\star}^{1 / 2} \equiv \mathcal{C}_{\star} \theta_{\rm E} \kappa_{\star}^{1 / 2}    
\end{equation*}

the approximate number of micro-caustics and the peak magnification can be expressed as

\begin{equation}
\label{eq:Number of microcaustic}
    N_\mathrm{c f} \simeq\left(\frac{2 \mathcal{C}_{\star}}{|\sin a|}\right)^{1 / 2}\left(\frac{\kappa_{\star}}{\kappa_{\rm c}}\right)^{3 / 4}\ , 
\end{equation}

\begin{equation}
\label{eq:Mupeak}
    \mu_\mathrm{peak} \simeq \frac{1}{\left|1-\kappa_{0}\right|}\left(\frac{D_{\rm s}}{R {|\bm d|}}\right)^{1 / 2}\left(\frac{\kappa_{\rm c}}{\kappa_{\star}}\right)^{3 / 4} \times\left\{\begin{array}{ll}
    1\ , & |\boldsymbol{\beta}|<s_{\rm w}\ ,  \\
    \left(s_{\rm w} /|\boldsymbol{\beta}|\right)^{3 / 8}\ , & |\boldsymbol{\beta}|>s_{\rm w}\ . 
    \end{array}\right.
\end{equation}

where $\mathcal{C}_{\star}=\left[\ln \left(3.05 N_{\mathrm{\star}}^{1 / 2}\right)\right]^{1 / 2}$, $R$ represents the radius of the source.
They suggested that the peak magnification remains approximately constant within the micro-caustic band and is proportional to $\left(\kappa_{\star}/{\kappa_{\rm c}}\right)^{3/4}$. 
Outside the micro-caustic band, as the distance from the caustic line increases, the peak magnification rapidly decreases following $\left(s_{\rm w} /|\boldsymbol{\beta}|\right)^{3 / 8}$.

By employing the ray-tracing method described in Section 3, we obtained a large number of source plane magnification maps and light curves, which allows us to compare their results of micro-caustic (peak) number density and peak magnification with our numerical statistics.
The solid black lines in Fig3$\rm {(b)}$, $\rm {(c)}$ and $\rm {(d)}$ represent the 5000th column and the adjacent two columns of light curves on the source plane in the vertical range of 900 to 1000. 
It can be observed that there are pronounced ``jitters" at the bottom of these light curves. These ``jitters" are caused by Poisson noise due to fluctuations in the number of rays and the presence of faint micro-caustics.
Therefore, it is crucial to determine whether the peaks on the light curves are caused by Poisson noise or by the crossing of micro-caustics, before quantifying the number of peaks and peak magnifications statistically. 
For ease of explanation, we will refer to them as ``false peaks" and ``true peaks," respectively.

\begin{figure}[H]
\center{\includegraphics [scale=0.68]{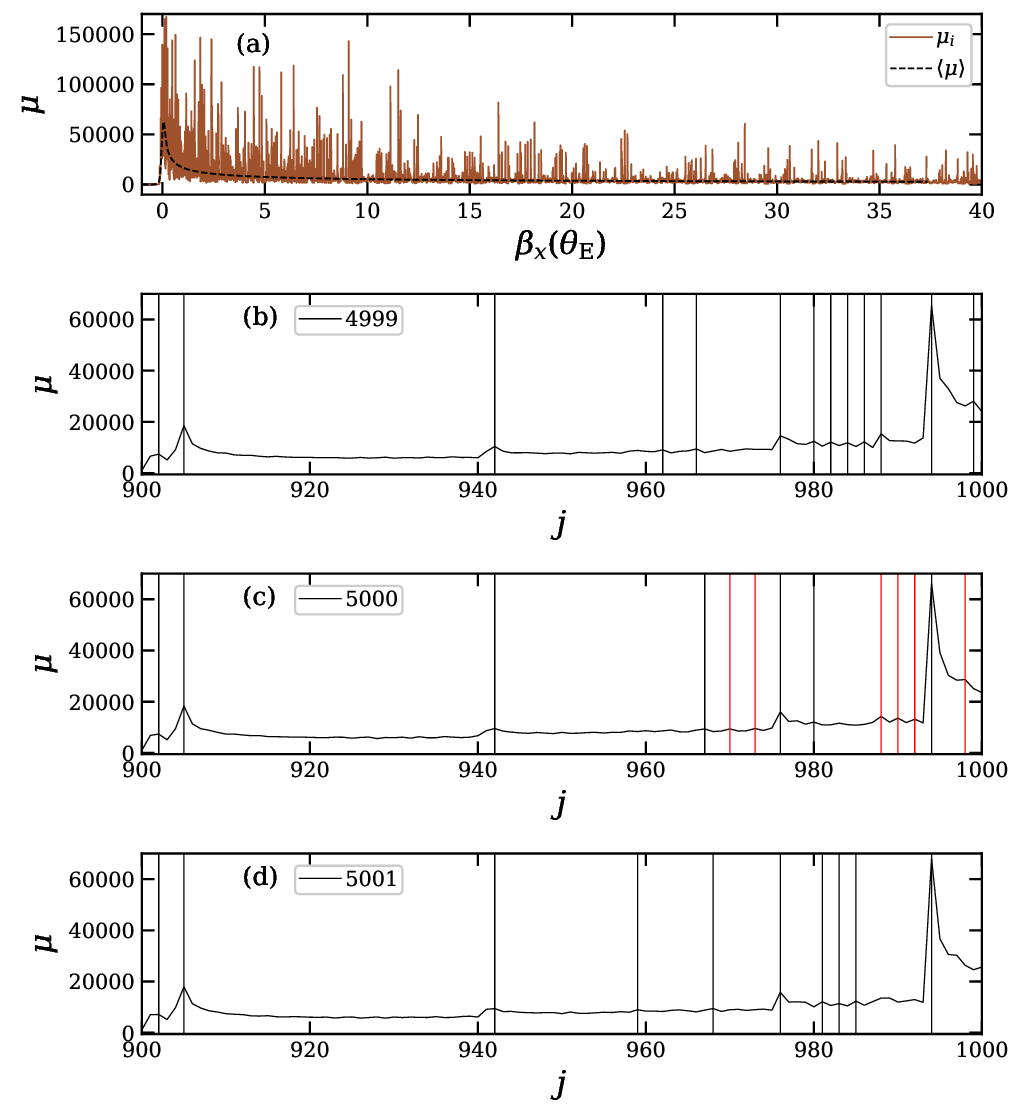}}
\caption{Light curves and example diagram of peak identification. The solid brown line in the figure ${\rm (a)}$ represents the change of the magnification in each source plane pixel as the background source moves vertically downward along column 5000 of the magnification map, where $\beta_{x}$ is the vertical position of the source plane, and the dashed black line represents the change of the average magnification along the vertical direction of the source plane. The black solid lines in the following three figures represents the changes of magnification  that the background source moving vertically downward along column 4999, 5000, and 5001 of the source plane. $j$ represents the vertical label of the source plane pixel. Both the black and red vertical lines indicate candidate peaks with confidence higher than $5\sigma_{ij}$, and red vertical lines mark candidate peaks excluded after cross validation in column 5000. }
\label{fig:lightcurve}
\end{figure}

Taking the peak identification of the amplification variations in the $i$-th column of the source plane as an example, if the light count $N_{ij}$ of the $(i, j)$-th pixel satisfies the conditions $N_{ij} > \mathrm{Max}\left[ N_{i(j-1)}, N_{i(j+1)}\right]$ and a relative peak height $h_{ij} > 5 \sigma_{ij}$, then the position of the $(i, j)$-th pixel is marked as a ``candidate peak", where $h_{ij}=N_{ij}-\langle N\rangle$, $\langle N\rangle=\left(N_{i(j-1)}+N_{ij}+N_{i(j+1)}\right)/3$, $\sigma_{ij}=\sqrt{\langle N\rangle}$.
By setting these two criteria, we identify the candidate peaks with a confidence level greater than $5\sigma_{ij}$.
Furthermore, we can further utilize the adjacent amplification curves on the left and right sides to eliminate false peaks. Since the caustic form continuous curves, there should be peaks present in the three neighboring pixels on both sides of a peak. This means that a candidate peak at position $(i, j)$ will be considered a true peak only if one of the three pixels on the left, $(i-1, j-1)$, $(i-1, j)$, and $(i-1, j+1)$, is marked as a candidate peak, and at the same time, one of the three pixels on the right, $(i+1, j-1)$, $(i+1, j)$, and $(i+1, j+1)$, is also marked as a candidate peak.

Fig.\ref{fig:lightcurve} (b), (c), and (d) illustrate the process of peak identification in the 5000th column of the source plane. 
The black and red vertical lines in the figure represent examples of candidate peaks identified in the first step from the 4999th, 5000th, and 5001st columns.
After the second step of cross-validation, the candidate peak marked in red on the 5000th column is excluded. 
The black marks in Fig.\ref{fig:lightcurve} (c) indicate the peaks that are ultimately identified.

\begin{figure}[H]
\center{\includegraphics [scale=0.55]{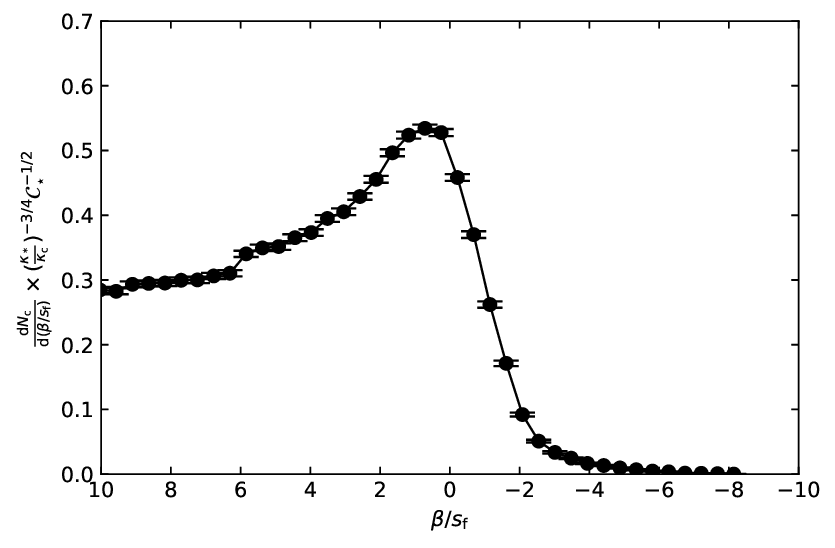}}
\caption{The relationship between  microcaustic density and the distance from source to macrocaustic . The black dots are the results of our statistics of 800 light curves, and the black error bar is a statistical error. For the convenience of display, the ordinate is the microcaustic density  scaled by the factor in the formula (\ref{eq:Number of microcaustic}), and the abscissa is the distance from source to macrocaustic scaled by $s_{\rm f}$. }
\label{fig:numpeak}
\end{figure}

We generate 80 source plane magnification map as described in section 4.1, and extract 800 light curves for exploring the statistical characteristics of peak magnifications.
On the NVIDIA Tesla V100S PCIe 32GB GPU, the average computation time for calculating a magnification map is approximately 7000s.
Fig.\ref{fig:numpeak} illustrates the variation of the normalized micro-caustic number density with the distance from the source to the caustic curve, as obtained from our statistics. 
Here, the micro-caustic number density is normalized using the factor given in equation (\ref{eq:Number of microcaustic}), while the distance from the source to the caustic curve is normalized with $s_{\rm f}$.
Comparing Fig.\ref{fig:numpeak} with the Fig.7 from Venumadhav et al \cite{V2017}, it can be seen that the normalized micro-caustic number density is approximately $0.3s_{\rm f}^{-1}$ at large distances from caustic, and there is a plateau in the micro-caustic number density at the peak of both figures.
Furthermore, there are two notable differences: (1) Our normalized micro-caustic number density at the peak, approximately $0.53 s_{\rm f}^{-1}$, is slightly lower than theirs, approximately $0.8 s_{\rm f}^{-1}$. They used Newton's iteration method to calculate the magnification, which theoretically allows them to identify all micro-caustics.
In our ray-tracing algorithm, however, due to the presence of some faint micro-caustics within high-magnification backgrounds, our peak identification method fails to recognize these weak micro-caustics, resulting in lower results compared to theirs. However, from another perspective, our results may be more consistent with the observational selection process.
(2) The width of the plateau in the micro-caustic density at its peak is different between our result and theirs. Our plateau width is approximately $2s_{\rm f}$, which is only half of theirs. Due to algorithm improvements, we are able to sample more extensively, resulting in lower noise, which leads to a narrower plateau structure in our statistical analysis.

\begin{figure}[H]
\center{\includegraphics [scale=0.6]{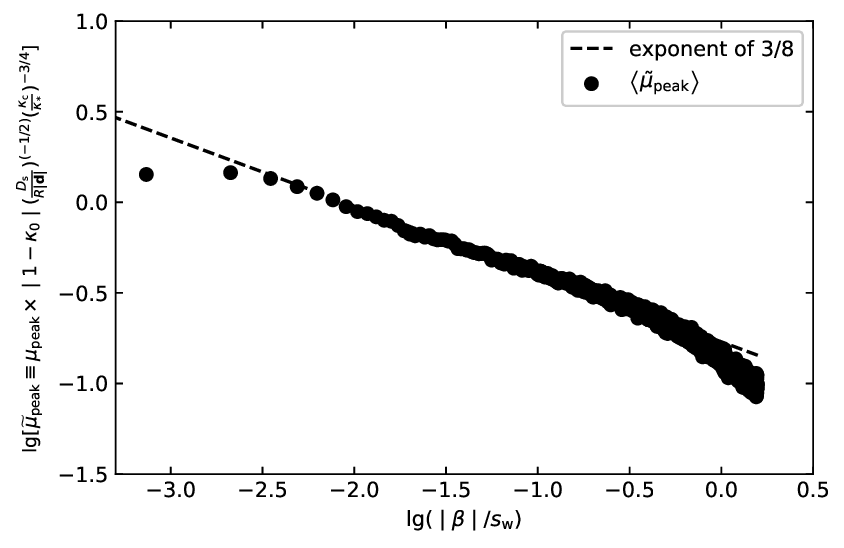}}
\caption{The relationship between the mean value of peak magnification and distance from source to macrocaustic. The black dashed line represents the trend of the peak magnification with distance as described by Eq.(\ref{eq:Mupeak}) with exponent of 3/8. For the convenience of display, the ordinate is the mean value of peak magnification scaled by the factor in the formula (\ref{eq:Mupeak}), and the abscissa is the distance from source to macrocaustic scaled by $s_{\rm w}$. Both abscissa and ordinate take logarithm based on 10. }
\label{fig:mupeak}
\end{figure}

Then, we investigated the peak magnifications on the light curves. 
The black dots in Fig.\ref{fig:mupeak} depict the variation of the reduced average peak magnification $\left \langle \tilde{\mu}_\mathrm{peak}\right \rangle$ with respect to the source-to-caustic distance in our simulated light curves. The black dashed line represents the power-law behavior with an exponent of 3/8 as given by Equation (\ref{eq:Mupeak}).
For brevity, the source-to-caustic distance is reduced with $s_{\rm w}$, while the average peak magnification is reduced with the factor given in Equation (\ref{eq:Mupeak}).
Compared to Figure 8 in Venumadhav et al \cite{V2017},  our results are more stable due to the improved sampling rate, at the same time, there are also some similarities and differences. 
(1) The reduced average peak height is approximately 0.1 at a distance of one $s_{\rm w}$.
(2) Within one $s_{\rm w}$, numerical results of Venumadhav et al \cite{V2017} remain almost constant, exhibiting a plateau behavior consistent with their theoretical estimates. However, the height of the plateau is only 1/4 of their theoretical prediction.
On the other hand, as the source approaches the caustic, our results continue to increase according to $\left(s_{\rm w} /|{\bm \beta}|\right)^{3/8}$ until they reach a constant value of 1 at $< 0.01 s_{\rm w}$.

By examining Fig.\ref{fig:lightcurve}${\rm (a)}$ in conjunction, we can investigate this difference. In this work, we have $s_{\rm w}=25.5\theta_{\rm E}$, and it is evident that the peak magnification is not a constant within one $s_{\rm w}$. As $\beta_{x} \to 0$, there is a noticeable brightening trend in the peak magnification, which provides further evidence for the accuracy of our statistical results.

\section{SUMMARY}

In this work, we have employed an efficient GPU-based ray tracing algorithm to study the microlensing effects near critical curves. On the NVIDIA Tesla V100S PCIe 32GB GPU, it takes approximately 7000 seconds to perform high-precision and high-resolution simulations that involve processing tens of thousands of microlenses and emitting light rays on the order of $10^{13}$. 
We have chosen a higher microlens surface mass density ($\kappa_{\star}=0.001$) than that in the study of Venumadhav et al \cite{V2017}, and we have applied this simulation to generate 800 light curves for investigating the statistical characteristics of microlensing effects near critical curves.

We found that near the caustic, the gravitational microlensing effect exhibits distinct microcaustic structures. 
These microcaustic curves overlap with each other, causing strong perturbations to the caustic and forming a prominent microcaustic band in its vicinity. 
Our statistical results show good consistency with the findings of Venumadhav et al \cite{V2017}, but with slight differences: (1) The reduced microcaustic density is approximately $0.3 s_{\rm f}^{-1}$ at large distances, and the density exhibits a plateau at its peak, with a width of approximately $2s_{\rm f}$ and a reduced height of approximately $0.53 s_{\rm f}^{-1}$. 
(2) As the source approaches the caustic, the peak magnification increases with $\left(s_{\rm w} /|\beta|\right)^{3 / 8}$, until it reaches a stable value within $\sim 0.01s_{\rm w}$. 
Considering that our statistical results depend on the peak identification method, which retains peaks with higher signal-to-noise ratios. 
It is more consistent with the actual observational selection process. We believe this could be one possible explanation for the aforementioned differences. 
Additionally, due to significant improvements in computational efficiency, we were able to conduct thorough sampling, resulting in lower noise levels and higher reliability in our statistical results.

\section*{Acknowledgements}
We thank Professor Guoliang Li for his guidance and support throughout this work.
We acknowledge the cosmology simulation database (CSD) in the National Basic Science Data Centre (NBSDC) and its funds the NBSDC-DB-10 (No. 2020000088)

\end{document}